\documentclass[conference,a4paper]{APSIPA2026}
\usepackage{amsmath}
\usepackage{graphicx}
\usepackage{multirow}
\usepackage{threeparttable}
\usepackage[
  backend=biber,
  style=ieee,
  giveninits=true,
  maxbibnames=5,
  minbibnames=1,
  doi=false,
  url=false,
  isbn=false,
  eprint=false
]{biblatex}
\usepackage{amssymb}

\usepackage{comment}
\usepackage{caption}
\usepackage{booktabs}
\usepackage[table,xcdraw]{xcolor}
\usepackage{makecell}
\usepackage{tikz}
\usepackage{pifont}
\usepackage{colortbl} 
\usepackage[table,xcdraw]{xcolor}
\usepackage[normalem]{ulem}
\useunder{\uline}{\ul}{}

\usepackage{geometry}
\usepackage{fancyhdr}

\fancypagestyle{firststyle}{
  \fancyhf{}
  \fancyhead[C]{2026 Asia Pacific Signal and Information Processing Association Annual Summit and Conference (APSIPA ASC)}
}

\begin{document}
\IEEEoverridecommandlockouts
\title{Contrastive Learning with Variational Regularization for Multi-Session EEG-to-Speech Decoding}

\author{
\authorblockN{
Tomoaki MIZUNO\authorrefmark{1}
Toru NAKASHIKA\authorrefmark{1}
}

\authorblockA{
\authorrefmark{1}
The University of Electro-Communications, Tokyo, Japan \\
E-mail: t.mizuno2301@uec.ac.jp 
}
\thanks{This work was partially supported by JSPS KAKENHI Grant Number JP24H00715 and by JST BOOST, Japan Grant Number JPMJBS2415. The authors thank the Artificial Intelligence eXploration Research Center (AIX) at The University of Electro-Communications for providing computational resources.}}

\maketitle
\thispagestyle{firststyle}
\pagestyle{empty}

\begin{abstract}
Reconstructing heard speech from non-invasive electroencephalography (EEG) is challenging due to a low signal-to-noise ratio (SNR) and inter-session variability.
While trial averaging improves the SNR, it is difficult to apply to continuous speech.
We instead use repeated EEG responses to the same stimulus across different sessions as positive pairs for contrastive learning, and introduce variational regularization that, combined with this contrastive objective, keeps the encoder representation space broad.
Experiments on a Japanese EEG dataset show that combining the session-invariant strategy with variational regularization improves the character error rate (CER) while maintaining mel-spectrogram reconstruction fidelity. Session probing confirms that the encoder representations achieve session-invariance.
\end{abstract}


\section{Introduction}
\label{sec:intro}
Speech synthesis from brain activity, which decodes speech content from neural activity and generates the corresponding speech, is one application of brain--machine interfaces (BMI).
While invasive approaches such as electrocorticography (ECoG) have achieved high-precision speech decoding~\cite{anumanchipalli2019,willett2023,metzger2023}, they require surgical electrode implantation.
In contrast, electroencephalography (EEG) is a non-invasive alternative that records electrical activity from the scalp.
Reconstructing heard continuous speech from EEG recorded during listening has been explored~\cite{lee2025,defossez2023,apsipa,dmf2mel,icassp2024eeg}. %
However, current EEG-based reconstruction accuracy falls far short of practical use.
EEG signals inherently have a low signal-to-noise ratio (SNR), and the noise that obscures the target response includes inter-session non-stationarity~\cite{krumpe2017} and trial-to-trial latency jitter, in which the latency of the brain response varies across trials~\cite{ouyang2016,ouyang2020}.
Trial averaging is a standard technique for improving SNR in auditory EEG research, but it targets event-related potentials (ERP) and presupposes that the temporal structure of the responses is aligned across trials.
For continuous speech spanning more than a few seconds, latency fluctuations accumulate along the time axis, and because averaging causes the target response components to cancel one another, this technique is difficult to apply directly.
In this work, to extract components common to multiple trials without assuming temporal alignment between trials, we use multiple EEG signals as semantically aligned positive pairs for contrastive learning. 
Prior EEG contrastive approaches construct positive pairs based on temporal proximity or data augmentation~\cite{bendr, EEGPT}, responses from different subjects to the same stimulus~\cite{clsster}, or segments from the same subject without explicit stimulus alignment~\cite{subjectcl2025}.
In contrast, in this paper we propose a session-invariant contrastive learning strategy designed to reduce the influence of session-specific variability in multi-session EEG recordings. The strategy encourages representations of EEG responses to the same stimulus to remain consistent across recording sessions, thereby promoting stimulus-related representations that are less dependent on session-specific factors.
However, with contrastive learning alone, the encoder representations do not span a sufficiently wide region for speech synthesis, and the reconstruction fidelity is degraded.
We therefore incorporate a variational regularization term. Combined with contrastive learning, it keeps the encoder representation space from collapsing into a narrow region.
This differs from the conventional use of variational regularization, where the latent variable is passed to the decoder and serves as the basis for reconstruction, as in variational autoencoders~\cite{vae,contrastvae} or an information bottleneck~\cite{vib}; in our architecture, the variational latent variable is used only for regularization and is not fed to the decoder.
Contrastive learning and variational regularization have complementary effects: neither alone provides consistent improvements, whereas their combination expands the representation space and recovers the degraded reconstruction fidelity.

\section{Related works}
\label{sec:relatedworks}
In \cite{apsipa}, a Transformer-based sequence-to-sequence model was adapted to take EEG as input to reconstruct continuous speech from auditory EEG, with cross-attention enabling EEG-speech alignment. The model employs an encoder to map EEG signals into hidden representations $\boldsymbol{h}=[\boldsymbol{h}_{1},\ldots,\boldsymbol{h}_{T}] \in \mathbb{R}^{d\times T}$ where $d$ denotes the hidden dimension and $T$ is the number of time frames. An autoregressive decoder to generate log-mel spectrograms, which are subsequently converted to waveforms by a neural vocoder. We adopt this model as our baseline. 
The SpREAD dataset~\cite{spread} is a multi-session, multi-speaker auditory EEG dataset in which participants listened to the same speech stimuli across sessions. Baseline experiments reported in \cite{spread} applied this framework to the dataset, pooling all sessions without explicitly addressing inter-session variability. In this work, we exploit the repeated structure of this dataset by introducing a contrastive learning objective that treats EEG responses to the same stimulus as positive pairs regardless of the recording session.

\section{Method}
\label{sec:method}
\begin{figure}[!t]
  \centering
  \includegraphics[width=0.9\linewidth]{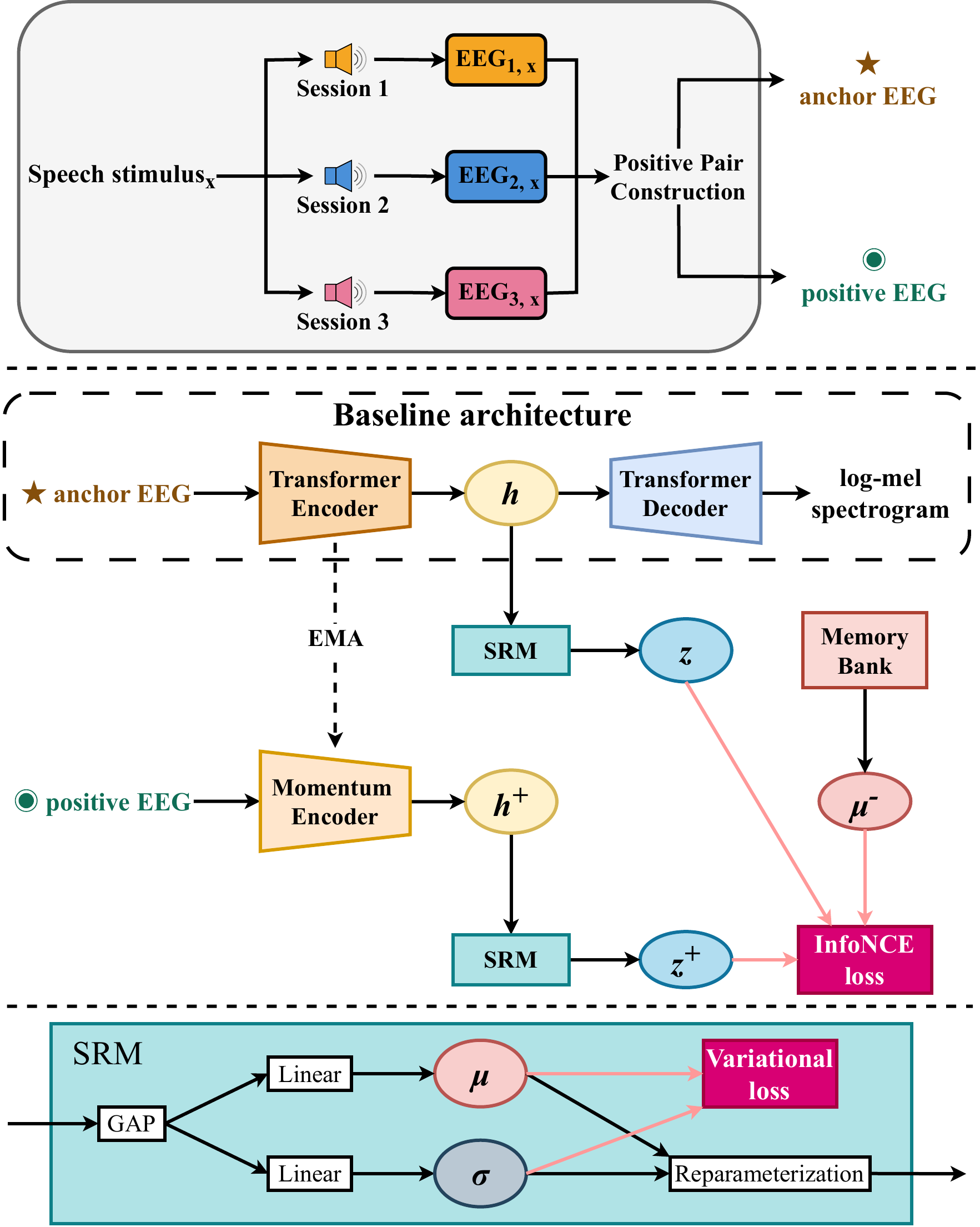}
  \caption{Overview of the proposed method. Top: Multi-session EEG data and positive pair construction. From repeated EEG recordings elicited by speech stimuli with the same linguistic content, we construct positive pairs from recordings of the same speech stimulus across different sessions. One recording is randomly selected as a anchor, and the remaining two serve as positive examples.
Middle: The architecture of the proposed model.
The Transformer Encoder maps the anchor EEG to $\boldsymbol{h}$, which is passed to the Transformer Decoder to predict log-mel spectrograms.
The SRM produces $\boldsymbol{z}$ via reparameterization; $\boldsymbol{z}$ is used only for the losses and is not passed to the decoder.
The momentum encoder, updated via the exponential moving average (EMA) of the encoder, generates $\boldsymbol{z}^{+}$ from the positive example.
InfoNCE is computed over $\boldsymbol{z}$, $\boldsymbol{z}^{+}$, and memory-bank negatives $\{\boldsymbol{\mu}^{-}\}$.
Bottom: Internal structure of the SRM: global average pooling (GAP) followed by linear layers predicting $\boldsymbol{\mu}$ and $\log\boldsymbol{\sigma}^{2}$; $\mathcal{L}_{\mathrm{KL}}$ regularizes the latent space.}
  \label{fig:proposed}
\end{figure}

Fig.~\ref{fig:proposed} illustrates the overall architecture, which builds upon the EEG-to-speech framework proposed in \cite{apsipa} (Section~\ref{sec:relatedworks}).
The baseline loss is
\begin{equation}
  \mathcal{L}_{\mathrm{baseline}} = \mathcal{L}_{\mathrm{L1}} + \alpha \mathcal{L}_{\mathrm{BCE}} + \beta \mathcal{L}_{\mathrm{att}},
  \label{eq:baseline_loss}
\end{equation}
where $\mathcal{L}_{\mathrm{L1}}$, $\mathcal{L}_{\mathrm{BCE}}$, and $\mathcal{L}_{\mathrm{att}}$ are the mel-spectrogram reconstruction, stop-token, and guided attention~\cite{guidedattn} losses, and $\alpha$, $\beta$ are weighting coefficients.

\subsection{Momentum Contrastive Learning with Variational Regularization}
\label{sec:moco}
We introduce momentum contrastive learning (MoCo) with a stochastic regularization module (SRM) into the EEG encoder to learn representations invariant to session variability. Leveraging multi-session recordings in which the same linguistic content is presented across sessions, we form positive pairs from EEG segments that share the same content, while SRM applies variational regularization to the encoder output.
In order to mitigate inter-session variability, we adopt momentum contrast (MoCo)~\cite{MoCo}, which maintains a momentum encoder (parameters $\phi$ updated via exponential moving average (EMA): $\phi \leftarrow m\phi + (1-m)\theta$, where $\theta$ denotes the online encoder parameters and $m$ is the momentum coefficient) and a memory bank of past representations.
Unlike typical setups where positive pairs are constructed via data augmentation, our pairs are drawn from separate EEG trials (Section~\ref{sec:positive_pairs}).

The contrastive objective operates on a latent variable $\boldsymbol{z}$ produced by the SRM applied to the encoder output.
Since contrastive pairs are separate trials with no temporal alignment, the SRM first computes $\bar{\boldsymbol{h}} = \tfrac{1}{T}\sum_{t} \boldsymbol{h}_{t} \in \mathbb{R}^{d}$ via global average pooling (GAP), then predicts a diagonal Gaussian $q_{\psi}(\boldsymbol{z}\mid \bar{\boldsymbol{h}}) = \mathcal{N}(\boldsymbol{\mu},\,\mathrm{diag}(\boldsymbol{\sigma}^2))$ from which $\boldsymbol{z}=\boldsymbol{\mu}+\boldsymbol{\sigma}\odot\boldsymbol{\epsilon}$ is sampled via the reparameterization trick from a Gaussian noise $\boldsymbol{\epsilon}\sim\mathcal{N}(\mathbf{0},\mathbf{I})$.
To prevent cluster degeneracy, the latent space is regularized by:
\begin{equation}
\mathcal{L}_{\mathrm{KL}}
= D_{\mathrm{KL}}\bigl(q_{\psi}(\boldsymbol{z}\mid \bar{\boldsymbol{h}})\,\Vert\,p(\boldsymbol{z})\bigr),
\quad
p(\boldsymbol{z})=\mathcal{N}(\mathbf{0},\mathbf{I}).
\label{eq:kl}
\end{equation}
The SRM weights are shared between the online and momentum paths.

At each training step, representations corresponding to positive pairs of the anchor are excluded from the memory bank queue, and the remaining $K$ entries serve as negatives $\{\boldsymbol{\mu^-}_i\}_{i=1}^{K}$. Each negative is the deterministic mean $\boldsymbol{\mu}$, while the anchor $\boldsymbol{z}$ and positive $\boldsymbol{z^+}$ are stochastic samples.
The InfoNCE objective is:
\begin{equation}
\mathcal{L}_{\mathrm{NCE}}
= -\log
\frac{s(\boldsymbol{z}, \boldsymbol{z^+})}
{s(\boldsymbol{z}, \boldsymbol{z^+})
+ \displaystyle\sum_{i=1}^{K} s(\boldsymbol{z}, \boldsymbol{\mu^-}_i)},
\label{eq:infonce}
\end{equation}
where $s(\mathbf{a},\mathbf{b})=\exp(\mathrm{sim}(\mathbf{a},\mathbf{b})/\tau)$, $\mathrm{sim}(\cdot,\cdot)$ denotes cosine similarity and $\tau$ is a temperature hyperparameter. 

Note that $\boldsymbol{z}$ is used only for the contrastive and variational objectives; the decoder cross-attends solely to the deterministic encoder output $\boldsymbol{h}$. This design differs from standard VAEs and their contrastive extensions such as ContrastVAE~\cite{contrastvae}, where  $\boldsymbol{z}$ is passed to the decoder as part of the generation pathway. In our architecture, variational regularization instead serves solely as an indirect regularizer on the encoder representations, discouraging the latent space from degenerating into a small number of clusters. 
\subsection{Cross-session positive pairs Construction}
\label{sec:positive_pairs}
This section defines the construction of positive pairs used in our contrastive learning framework to reduce session-induced variability. We focus on multi-session EEG data recorded from a single subject over multiple days. Since the same set of stimuli is presented in each session, EEG recordings corresponding to the same stimulus but obtained from different sessions of the same subject are treated as positive pairs.
This design aims to preserve EEG components associated with stimulus content while reducing the dependence of learned representations on session-specific factors, such as recording day, electrode placement, and changes in the subject’s physiological or cognitive state. \cite{clsster} used EEG responses to the same stimulus from different individuals as positive pairs to learn shared spatiotemporal representations across subjects. In contrast, our objective is not to learn representations shared across individuals, but to obtain representations that are invariant across sessions within a single subject.

\subsection{Training objective}

The overall training objective combines the baseline losses with the two proposed terms:
\begin{equation}
  \mathcal{L} = \mathcal{L}_{Baseline} + \gamma \mathcal{L}_{\mathrm{NCE}} + \delta \mathcal{L}_{\mathrm{KL}},
  \label{eq:total_loss}
\end{equation}
where $\mathcal{L}_{Baseline}$ is defined in Eq.~\eqref{eq:baseline_loss},
$\mathcal{L}_{\mathrm{NCE}}$ is the InfoNCE loss (Eq.~\eqref{eq:infonce}),
$\mathcal{L}_{\mathrm{KL}}$ is the KL divergence term (Eq.~\eqref{eq:kl}),
and $\gamma$, $\delta$ are weighting coefficients.

\section{Experiment}
\label{sec:experiments}
\begin{table*}[t]
\centering
\caption{Results on the evaluation set. \textbf{Bold}: significantly better than the baseline; \underline{underline}: significantly worse than the baseline (Holm--Bonferroni corrected, $\alpha=0.05$). CER: lower is better; SECS, PCC (dimensionless, $[-1,1]$) and SPA~(\%): higher is better. Values are mean~$\pm$~standard deviation over utterances, except SPA, which is reported as the mean~$\pm$~standard error.The proposed method is highlighted in gray.}
\label{tab:results}
\begin{tabular}{ccccccc}
\toprule
Tag                                & $\mathcal{L}_{\mathrm{NCE}}$ & $\mathcal{L}_{\mathrm{KL}}$ & CER~$\downarrow$     & SECS~$\uparrow$ & SPA~(\%)~$\uparrow$ & PCC~$\uparrow$    \\ \hline
Baseline        &     -                         &     -                        & 0.968±0.121          & 0.190±0.160     & 5.32±1.08           & 0.282±0.150       \\
CL alone                            & \checkmark                   &     -                        & \textbf{0.956±0.095} & 0.173±0.158     & 4.86±1.03           & {\ul 0.262±0.143} \\
VR alone                            &    -                          & \checkmark                  & 0.974±0.113          & 0.193±0.163     & 5.79±1.12           & 0.277±0.146       \\
\rowcolor{gray!25} Proposed method & \checkmark                   & \checkmark                  & \textbf{0.948±0.097} & 0.176±0.151     & 7.18±1.24           & 0.274±0.155       \\ \bottomrule
\end{tabular}
\end{table*}

\subsection{Dataset}
We use the SpREAD dataset
~\cite{spread}, recorded from a single participant listening to 1,353 Japanese utterances spoken by 18 speakers (9 male, 9 female) from the ASJ corpus~\cite{asj}.
Among the 1,353 utterances, 50 per speaker are shared across all 18 speakers, and the remaining 25--27 are unique to each speaker. 
Although the use of a single participant limits the generalizability of our findings, this setup allows us to focus on inter-session variability without confounding subject differences.
Each utterance was presented three times on separate days, yielding three sessions per stimulus.
The dataset comprises 45 recording sessions conducted over 9 days (5 sessions per day).
This repeated-measurement design enables the construction of cross-session positive pairs (Section~\ref{sec:positive_pairs}). 
EEG was recorded with a 64-channel Biosemi ActiveTwo system (Biosemi) at 2,048~Hz, preprocessed by downsampling to 1,024~Hz, bandpass filtering at 1--40~Hz, ocular artifact removal, and final downsampling to 512~Hz. 
Target speech was parameterized as 80-dimensional log-mel spectrograms at 16~kHz.
The data were split into training (3{,}195), development (432), and evaluation (432) samples, with all three recordings of the same text assigned to the same split.

\subsection{Experimental setup}

We evaluate six configurations in total.
Table~\ref{tab:results} summarizes the loss composition and evaluation results for the four main conditions.
Table~\ref{tab:ablation} presents two additional ablation conditions that isolate the individual contributions of contrastive learning (CL) and variational regularization (VR), along with two main conditions reproduced for ease of comparison.
The baseline uses only $\mathcal{L}_{\mathrm{Baseline}}$.
Building on this baseline, we further evaluate momentum contrastive learning ($\mathcal{L}_{\mathrm{NCE}}$) using the positive-pair strategies defined in Section~\ref{sec:positive_pairs}, variational regularization ($\mathcal{L}_{\mathrm{KL}}$), and their combination.
In the proposed method, $\boldsymbol{z}$ serves only as the target of variational regularization and the contrastive objective, and is not passed to the decoder. This differs from prior latent-variable approaches~\cite{vae,contrastvae,vib}, in which $\boldsymbol{z}$ is fed to the decoder as part of its input.
To examine whether explicitly providing $\boldsymbol{z}$ to the decoder yields further improvement, we additionally test a variant in which $\boldsymbol{z}$ is prepended as a single token to the temporal axis of $\boldsymbol{h}$ before decoding. This decoder input is denoted as $\boldsymbol{z}{+}\boldsymbol{h}$ in Table~\ref{tab:ablation}.
When variational regularization is disabled, the SRM is replaced by GAP alone, and $\bar{\boldsymbol{h}}$ is used directly in place of $\boldsymbol{z}$ for the contrastive objective.

Our model is based on the baseline described in Section~\ref{sec:relatedworks}, using the Voice Transformer Network (VTN)~\cite{VTN}\footnote{https://github.com/unilight/seq2seq-vc/tree/main/egs/arctic/vc1} with the same hyperparameter settings, except that we set the input dimensionality to 64 and use a batch size of 8.
The baseline loss weights were set to $\alpha = 10$ and $\beta = 1$.
The weighting coefficients for the proposed losses were set to $\gamma = 0.5$ and $\delta = 1$.
The temperature parameter was $\tau = 0.07$.
The momentum coefficient for the EMA update was $m = 0.999$, and the memory bank queue size was $4{,}096$.

Synthesized speech waveforms were generated from the predicted log-mel spectrograms using a HiFi-GAN~\cite{hifigan} vocoder~\footnote{https://github.com/jik876/hifi-gan} trained on the ground-truth speech of the same dataset.

\subsection{Evaluation metrics}
\label{sec:eval}
We evaluate the reconstructed speech from four aspects: linguistic fidelity, measured by the character error rate (CER); speaker identity, measured by the speaker encoder cosine similarity (SECS) and the speaker probing accuracy (SPA); acoustic reconstruction fidelity, measured by the Pearson correlation coefficient (PCC) of the mel-spectrogram; and session-invariance, measured by the session probing accuracy.

\noindent\textbf{Character Error Rate (CER).}\quad CER is the grapheme-level edit distance between transcriptions of generated and reference speech, $(S+D+I)/N$, where $S$, $D$, $I$, and $N$ are substitutions, deletions, insertions, and reference length. Both ground-truth and generated speech are transcribed using a pre-trained Japanese ASR model~\footnote{https://huggingface.co/AndrewMcDowell/wav2vec2-xls-r-1b-japanese-hiragana-katakana}. Text normalization is applied before computing CER. Lower is better.

\noindent\textbf{Speaker Encoder Cosine Similarity (SECS).}\quad SECS measures speaker similarity in a learned embedding space. We extract 192-dimensional speaker embeddings using a pre-trained ECAPA-TDNN~\cite{ECAPA} model\footnote{\url{https://huggingface.co/speechbrain/spkrec-ecapa-voxceleb}} and compute the cosine similarity between each generated utterance embedding and the corresponding ground-truth speaker centroid, defined as the mean embedding of that speaker's training-set ground-truth utterances. The 18 speakers are shared between training and evaluation. Higher is better.

\noindent\textbf{Speaker Probing Accuracy (SPA).}\quad SPA measures how much speaker identity is linearly decodable from the encoder latent representation $\boldsymbol{h}$. We train an SVM with a linear kernel, $C{=}1.0$, and one-vs-rest classification on utterance-averaged, $L_2$-normalized 384-dimensional encoder hidden states for 18-speaker classification. The classifier is trained on 3{,}195 training utterances and evaluated on 432 development utterances, with chance level $1/18\approx5.56\%$. Higher is better.

\noindent\textbf{Session Probing Accuracy.}\quad Session probing accuracy measures session information in $\boldsymbol{h}$ using the same linear SVM probing protocol as SPA, but with 45-class session labels. The classifier is trained on the same 3{,}195 training utterances and evaluated on 432 development utterances, with chance level $1/45 \approx 2.2\%$. Lower is better.

\noindent\textbf{Mel-spectrogram Pearson Correlation Coefficient (PCC).}\quad PCC measures acoustic reconstruction fidelity as the average Pearson correlation, over the 80 mel bands, between generated and ground-truth mel-spectrograms along the time axis. The mel-spectrograms are taken directly from the training pipeline before vocoder synthesis, using the standard 80-band configuration adopted in modern neural vocoders such as HiFi-GAN. Higher is better. PCC is commonly used for evaluating mel-spectrogram reconstruction from EEG of listened speech~\cite{icassp2024eeg,dmf2mel,lee2025}.

\noindent\textbf{Statistical testing.}\quad Each condition is compared against the baseline. For CER, SECS, and PCC, we use two-sided Wilcoxon signed-rank tests on utterance-level paired samples. For SPA and session probing accuracy, we use McNemar's exact test on utterance-level paired correct/incorrect predictions. For each metric, $p$-values are Holm--Bonferroni corrected with $\alpha=0.05$ within each group. Continuous metrics are reported as mean~$\pm$~standard deviation over utterances, and binary outcomes as $\hat{p}\pm\mathrm{SE}$ with $\mathrm{SE}=\sqrt{\hat{p}(1-\hat{p})/N}$. For session probing accuracy, we also use a two-sided binomial test against chance, $1/45 \approx 2.2\%$, to determine whether each condition reaches the floor effect. For ablations, the same paired tests are applied between the two variants of each base architecture, namely CL with VR and VR alone. Since each base architecture is an independent ablation with a single paired comparison per metric, no multiple-comparison correction is required.

\section{Result and Discussion}
\begin{table*}[t]
\centering
\caption{Results on the ablation set. For each group, the experiments in the upper row are without our proposed $\boldsymbol{z}$ as decoder input, while the experiments in the lower row with "+ prepending $\boldsymbol{z}$ to $\boldsymbol{h}$" tags shows results when $\boldsymbol{z}$ was concatenated with $\boldsymbol{h}$ before being fed into the decoder. For each of the experiments within each group, we conducted tests for each evaluation metric, but found no significant differences.The proposed method is highlighted in gray.}
\label{tab:ablation}
  \begin{tabular}{cccccc}
    \toprule
    Tag & Decoder input
      & CER $\downarrow$ & SECS $\uparrow$
      & SPA (\%) $\uparrow$ & PCC $\uparrow$ \\
    \hline
    \multicolumn{6}{c}{CL + VR} \\
    \rowcolor{gray!25}Proposed method
      & $\boldsymbol{h}$
      & $0.948{\pm}0.097$ & $0.176{\pm}0.151$
      & $7.18{\pm}1.24$   & $0.274{\pm}0.155$ \\
    \quad + prepending $\boldsymbol{z}$ to $\boldsymbol{h}$
      & $\boldsymbol{z}+\boldsymbol{h}$
      & $0.951{\pm}0.091$ & $0.174{\pm}0.148$
      & $6.02{\pm}1.14$   & $0.275{\pm}0.148$ \\
    \hline
    \multicolumn{6}{c}{VR alone} \\
    VR alone
      & $\boldsymbol{h}$
      & $0.974{\pm}0.113$ & $0.193{\pm}0.163$
      & $5.79{\pm}1.12$   & $0.277{\pm}0.146$ \\
    \quad + prepending $\boldsymbol{z}$ to $\boldsymbol{h}$
      & $\boldsymbol{z}+\boldsymbol{h}$
      & $0.971{\pm}0.119$ & $0.185{\pm}0.149$
      & $6.71{\pm}1.20$   & $0.275{\pm}0.140$ \\
\bottomrule
  \end{tabular}
\end{table*}

\begin{table}[t]
\centering
\caption{Session probing accuracy (\%).
Chance level is $1/45 \approx 2.2\%$; lower is better. $\dagger$:~indistinguishable from chance (two-sided binomial test, $p > 0.05$). The proposed method is highlighted in gray.}
\label{tab:session_probe}
\begin{tabular}{lccc}
\toprule
 & w/o VR & w/ VR & \makecell{w/ VR \\ +prepending $\boldsymbol{z}$ to $\boldsymbol{h}$} \\
\midrule
w/o CL & 25.2$\pm$2.09 & 24.8$\pm$2.08 & 23.2$\pm$2.03 \\
w/ CL & 2.78$\pm$0.79$\dagger$ & 
\cellcolor{gray!25}2.08$\pm$0.69$\dagger$ & 3.01$\pm$0.82$\dagger$ \\
\bottomrule
\end{tabular}
\end{table}

\begin{figure}[t]
  \centering
  \includegraphics[width=\linewidth]{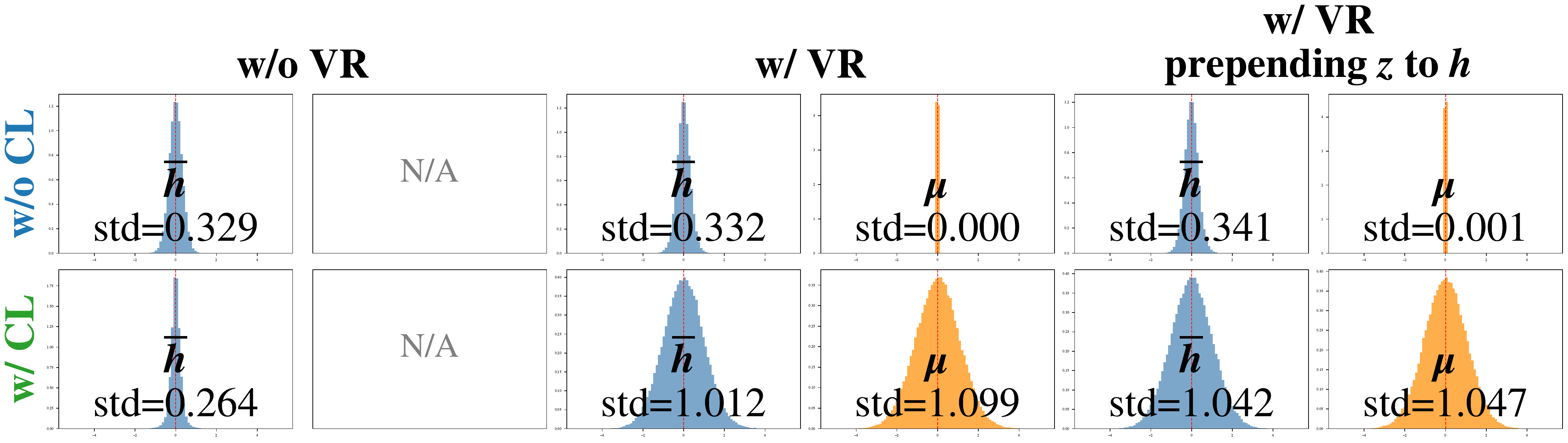}
 
  \caption{Value distributions of the encoder summary vector $\bar{\boldsymbol{h}}$ (blue) and the SRM posterior mean $\boldsymbol{\mu}$ (orange) across test samples.The proposed method corresponds to w/ CL + w/ VR.}
  \label{fig:dist}
\end{figure}

Table~\ref{tab:results} shows results for CER (linguistic fidelity), SECS and SPA (speaker identity), and PCC (mel-spectrogram reconstruction fidelity), Figure~\ref{fig:dist} shows the value distributions of $\bar{\boldsymbol{h}}$ and $\boldsymbol{\mu}$, and Table~\ref{tab:session_probe} additionally quantifies the session-invariance of the encoder output via the session probing accuracy described in Section~\ref{sec:eval}.
 
\noindent\textbf{Contrastive learning alone.}\quad
Adding the contrastive loss to the baseline significantly improved CER but significantly degraded PCC, while SPA showed no significant difference.
The session probing accuracy dropped to chance level and was statistically indistinguishable from chance by a two-sided binomial test, confirming that contrastive learning achieved session-invariance at the encoder level.
Despite this successful session separation, the CER improvement accompanied losses in PCC without variational regularization, constituting a trade-off between linguistic encoding and acoustic fidelity.
 
\noindent\textbf{Combining contrastive learning with variational regularization.}\quad
Variational regularization on its own yielded no significant improvement in CER or speaker identity, and its PCC remained at the baseline level.
Its session probing accuracy remained at the baseline level and significantly exceeded chance by a binomial test, indicating that variational regularization alone did not reduce session dependence.
Adding variational regularization to contrastive learning retained the CER improvement and significantly improved PCC in the paired comparison ($p=0.046$), restoring it to the baseline level and eliminating the degradation observed under contrastive learning alone.
Although the difference in SPA did not reach significance ($p=0.087$), a medium effect size (Cohen's $g=0.179$) indicated a trend toward improvement.
The session probing accuracy of the proposed method also dropped to chance level (indistinguishable from chance, $p=1.00$), confirming that the addition of variational regularization preserved the session-invariance achieved by contrastive learning.
 
To understand why the combination recovers the PCC degradation observed under contrastive learning alone, we examine the latent representations under each condition.
As shown in Figure~\ref{fig:dist}, variational regularization alone caused complete posterior collapse, reproducing the failure mode reported for CR-VAE~\cite{cr-vae}; the collapse was confined to the VAE projection and did not flatten the encoder itself, as the encoder summary vector $\bar{\boldsymbol{h}}$ retained std $=0.332$.
Contrastive learning alone also yielded a narrow $\bar{\boldsymbol{h}}$ distribution (std $=0.264$).
When the two were combined, $\boldsymbol{\mu}$ recovered a near-Gaussian spread (std $=1.099$) and $\bar{\boldsymbol{h}}$ widened accordingly (std $=1.012$).
Since $\boldsymbol{\mu}$ is a linear projection of $\bar{\boldsymbol{h}}$, the variational-regularization gradient propagates directly to $\bar{\boldsymbol{h}}$ and increases its dispersion.
By driving the latent distribution toward $\mathcal{N}(\mathbf{0}, \mathbf{I})$, variational regularization prevents the representation from concentrating into a narrow region, addressing both the limited dispersion of contrastive learning alone and the posterior collapse of variational regularization alone.
The recovery of dispersion coincided with the recovery of PCC, and the SPA trend was likewise consistent with the widening of $\bar{\boldsymbol{h}}$.
Session probing confirmed that the session-invariance observed under contrastive learning alone was preserved when variational regularization was added.
Overall, these results show that the combination maintains session-invariance while expanding the representation space, yielding complementary benefits: preserved CER improvement, restored PCC, and a positive trend in speaker identity. This validates that the proposed contrastive learning strategy successfully reduces the influence of session-specific variability in multi-session EEG recordings.
 
\noindent\textbf{Effect of prepending $\boldsymbol{z}$ to $\boldsymbol{h}$.}\quad
Table~\ref{tab:ablation} presents the results of the ablation study.
Our architecture uses $\boldsymbol{z}$ only for the regularization and contrastive losses and does not pass it to the decoder.
To validate this design choice, we also evaluated a variant that prepends $\boldsymbol{z}$ to $\boldsymbol{h}$ before the decoder.
For variational regularization alone, the variant with prepending showed no significant difference from the variant without prepending on any of the four metrics, and all effect sizes were negligible.
For contrastive learning with variational regularization, the variant with prepending also showed no significant difference on any of the four metrics (all $p > 0.05$); effect sizes were negligible for CER, SECS, and PCC, and small only for SPA (Cohen's $g=0.081$), where the variant without prepending was numerically higher ($7.18$ vs.\ $6.02$).
Moreover, the value distributions in Figure~\ref{fig:dist} and the session probing accuracy were also consistent between the two variants, with both remaining indistinguishable from chance, confirming that prepending $\boldsymbol{z}$ to $\boldsymbol{h}$ did not alter the learned representations.
Since both variants achieved comparable improvements over the baseline, the simpler design without prepending is preferable as prepending $\boldsymbol{z}$ only increases the attention computation in the decoder without additional gain. This supports our architecture where $\boldsymbol{z}$ is used solely as a regularization signal.
 
\section{Conclusions}

We proposed a contrastive learning framework with indirect variational regularization for EEG-to-speech decoding. Cross-session positive pairs promote session-invariant representations, while variational regularization encourages the encoder representations to span a sufficiently broad region for speech synthesis. Neither component alone yielded consistent improvements: contrastive learning alone improved CER but degraded PCC, whereas variational regularization alone produced no clear gains. Their combination preserved the CER improvement, recovered PCC, and expanded the encoder representation space, suggesting complementary roles. Session probing showed that all contrastive-learning conditions reached the floor effect, with accuracy statistically indistinguishable from chance, supporting reduced session-specific information in the encoder representation. Feeding $\boldsymbol{z}$ to the decoder yielded no benefit, supporting its use solely as a regularization signal. These results suggest that cross-session contrastive learning, which does not assume temporal alignment across trials, combined with variational regularization helps extract EEG representations relevant to listened speech.

\printbibliography

\end{document}